\documentclass[aps,prl,twocolumn,superscriptaddress,amsmath,amssymb,showpacs]{revtex4-1}
 \usepackage{graphicx,graphics,color}
\usepackage{amsmath}
\usepackage{amssymb}
\usepackage{amsfonts}
\usepackage{amsfonts}
\usepackage{epstopdf}
\usepackage{bm}
\usepackage{times,xspace}
\usepackage{color}
\usepackage[utf8]{inputenc}
\usepackage[normalem]{ulem}
\usepackage{xcolor}
\usepackage{multirow}

\newcommand{\comment}[1]{}

\begin{document}

\title{Oligomodal mechanical metamaterials}

\author{Aleksi Bossart}
\thanks{These two authors contributed equally}
\affiliation{Institute of Physics, University of Amsterdam,  Science Park 904, 1098 XH, Amsterdam, the Netherlands}
\affiliation{Laboratory of Wave Engineering, École polytechnique fédérale de Lausanne, 1015 Lausanne, Switzerland}
\author{David M.J. Dykstra}
\thanks{These two authors contributed equally}
\affiliation{Institute of Physics, University of Amsterdam,  Science Park 904, 1098 XH, Amsterdam, the Netherlands}
\author{Jop van der Laan}
\affiliation{Institute of Physics, University of Amsterdam,  Science Park 904, 1098 XH, Amsterdam, the Netherlands}
\author{Corentin Coulais}
\email[E-mail: ]{coulais@uva.nl}
\affiliation{Institute of Physics, University of Amsterdam,  Science Park 904, 1098 XH, Amsterdam, the Netherlands}


\begin{abstract} 
Mechanical metamaterials are artifical composites that exhibit a wide range of advanced functionalities~\cite{Bertoldi_NatRevMat} such as negative Poisson's ratio~\cite{Bertoldi_AdvMat2010,Shim_PNAS2012}, shape-shifting~\cite{Shim_PNAS2012,Milton_JMPS2013,Cho_PNAS2014,Coulais_Nature2016,Frenzel_Science2017,Coulais_Nature2018,Maha_NatMat2019,Shaw_NatCom2019,Bessa_AdvMat2019}, topological protection ~\cite{KaneLubensky_NatPhys2014,Paulose_NatPhys2015,Rockling_PRL2016,Meeussen_NatPhys2020}, multistability~\cite{Rafsanjani_AdvMat2015,Silverberg_NatMat2015,Stern_NatCom2018,Jin_PNAS2020}, and enhanced energy dissipation~\cite{Frenzel_AdvMat2016,Dykstra_JAM2019}. To date, most metamaterials have a single property~\cite{Bertoldi_AdvMat2010,Shim_PNAS2012,Coulais_Nature2016,Frenzel_Science2017,KaneLubensky_NatPhys2014,Paulose_NatPhys2015,Rockling_PRL2016}, e.g. a single shape change, or are pluripotent~\cite{Cho_PNAS2014,Coulais_Nature2018,Liarte_arXiv2020}, \emph{i.e.} they can have many different responses, but require complex actuation protocols. 
Here, we introduce a novel class of oligomodal metamaterials that encode a few distinct properties that can be selectively controlled under uniaxial compression. 
In particular, we realise a metamaterial that has a negative (positive) Poisson's ratio for low (high) compression rate. The ability of our oligomodal metamaterials to host multiple mechanical responses within a single structure makes them an early example of multi-functional matter and paves the way towards robust and adaptable devices.
\end{abstract}

\maketitle

\begin{figure*}[t!]
\centering 
\includegraphics[width=1.5\columnwidth]{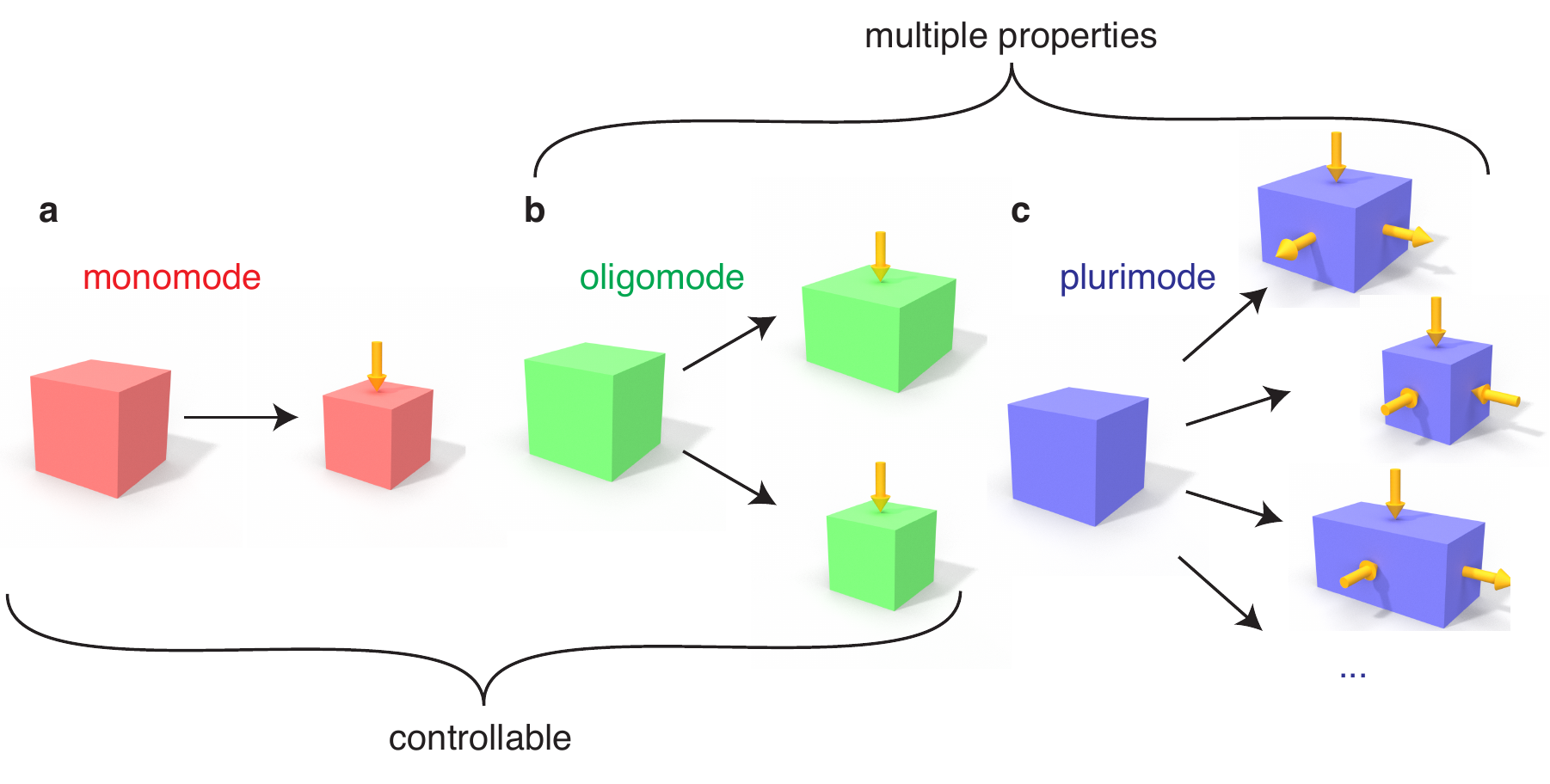}
\caption{\textbf{Oligofunctional materials} (a) Monomodal materials have only one property, that can be obtained via a simple actuation protocol. (b) Oligofunctional materials have a few distinct properties, that can be selected with a simple actuation protocol, e.g. uniaxial compression. (c) Plurimodal materials are kinematically able to host a large number of properties but require complex actuation protocols, e.g. multiaxial loading.}
\label{fig1}
\end{figure*}

The functionalities of mechanical metamaterials are often rooted in soft deformation modes, which typically mimic a mechanism with desirable kinematic properties~\cite{Bertoldi_NatRevMat}. So far, such mechanism-based metamaterials are either monomodal (Fig. \ref{fig1}a) or plurimodal (Fig. \ref{fig1}c). On the one hand, monomodal metamaterials feature a single soft deformation mode and a single functionality~\cite{Bertoldi_AdvMat2010,Shim_PNAS2012,Coulais_Nature2016,Frenzel_Science2017,KaneLubensky_NatPhys2014,Paulose_NatPhys2015,Rockling_PRL2016,Maha_NatMat2019}, which is typically robust and easy to control with a simple actuation protocol, e.g. uniaxial compression. On the other hand, plurimodal metamaterials feature a large number of soft deformation modes increasing with system size~\cite{Cho_PNAS2014,Coulais_Nature2018,Liarte_arXiv2020}, which offer multiple possible functionalities, but that are hard to control, e.g. they require complex actuation protocols for successful execution. The challenge we address here is whether it is possible to find a middle ground between monomodal and plurimodal metamaterials. In order words, can we design and create oligomodal metamaterials (Fig. \ref{fig1}b), \emph{i.e.} metamaterials that have more than one mode, but not an uncontrollably large number? Could these modes lead to interesting mechanical properties? Could these metamaterials be actuated in a robust fashion with a simple actuation protocol?

\begin{figure*}[t!]
\centering
 \hspace{0in} \includegraphics[width=1.0\columnwidth]{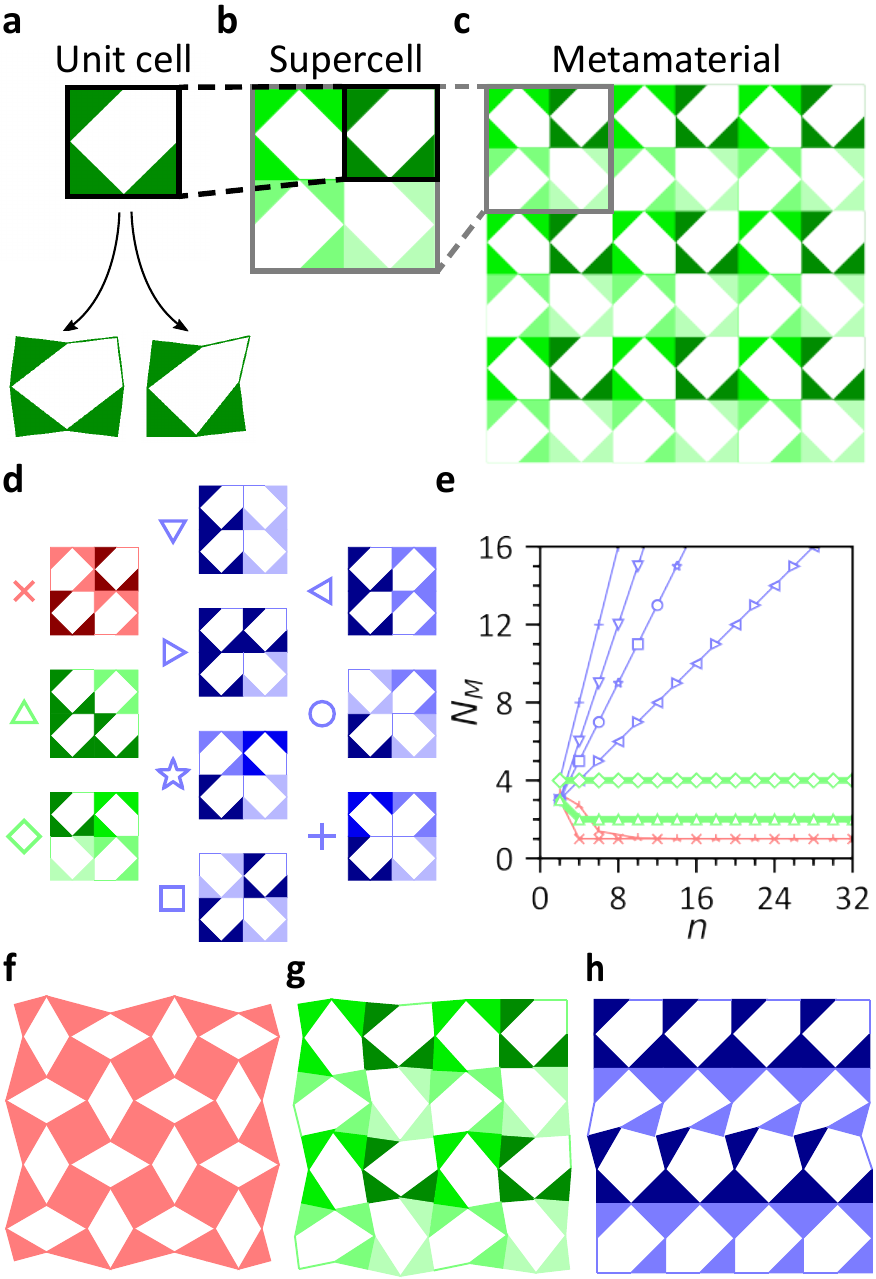}  \\
\caption{\textbf{Combinatorial design space} (a) Bimodal unit cell that has two soft deformation modes and that can be stacked in four orientations to create $2\times2$ supercells (b). (c) Periodic $6\times6$ metamaterial made from stacking a $2\times2$ supercell. (d) All mechanically non-equivalent $2\times2$ supercells, featuring monomodal (red), oligomodal (green) and plurimodal (blue) mode scalings. (e) Mode scaling, $N_M$, with the tiling's side length, $n$, for all ten tilings in (d) and the average of random tilings (red triangular crosses). (f) Rotating-squares mechanism of monomodal tilings. (g) Spatially extended mode typical of oligomodal tilings. (h) Line localised mode typical of plurimodal tilings.}
\label{fig2}
\end{figure*}

To first design oligomodal architectures, we turn towards combinatorial metamaterials, which is a particularly fruitful paradigm for the design of advanced mechanical functionalities~\cite{Coulais_Nature2016,Dieleman_NatPhys2019,Meeussen_NatPhys2020}. In combinatorial metamaterials, the structural complexity is reduced to a discrete design space, typically by controlling the orientation of the constitutive unit cells. Such discreteness makes the design space much easier to explore and has recently been leveraged to create non-periodic metamaterials with shape-changing~\cite{Coulais_Nature2016,Dieleman_NatPhys2019}, textured functionality~\cite{Coulais_Nature2016} and topological properties~\cite{Meeussen_NatPhys2020}, yet only for single mode metamamaterials so far. Here, we generalise combinatorial strategies to metamaterials with more than one mode. To this end, we introduce a unit cell with two soft deformation modes (Fig. \ref{fig2}a). A central ingredient in combinatorial design is that the unit cells have less symmetries than that of the lattice they live on. In the present case, the unit cell does not have any planar rotational symmetry. Thereby, it can be tiled in four different orientations on a square lattice (Fig. \ref{fig2}b), in turn allowing us to construct a wide range of combinatorial metamaterials.

We first analyse the combinatorial landscape of the metamaterials and notice that the design space is extremely large, \textit{i.e.} for a $n\times n$ tiling, there are $4^{n^2}$ possible configurations. It is thus impossible to sample all the possible configurations even numerically. Therefore, we restrict our attention to a subset of configurations. Namely, we will focus on periodic tilings of $2\times 2$ supercells (Fig. \ref{fig2}c), which is a very small fraction of the design space---we discuss an example of quasiperiodic tilings in the Appendix. Any choice of four orientations defines a $2\times 2$ supercell and thus a tiling; hence there are 256 possible configurations. By using the isometries of the square, we can reduce these to only ten mechanically non-equivalent configurations (Fig. \ref{fig2}d). A numerical study of the kinematics (Appendix) then reveals a rich spectrum of soft deformation modes. This spectrum consists of the three classes of metamaterials defined in Fig.~\ref{fig1}. For system sizes $n>4$, monomodal tilings (Fig \ref{fig2}de, red) have a single mode; oligomodal tilings have a constant mode number larger than one (Fig \ref{fig2}de, green); and plurimodal tilings have a number of modes that increases linearly with $n$ (Fig \ref{fig2}de, blue). All three classes of tilings share a common mode, which is the rotating-squares mechanism~\cite{grima_auxetic}, where the unit cells have the same deformation mode in a spatially alternating manner (Fig.~\ref{fig2}f). Monomodal tilings only have this mode while oligomodal and plurimodal tilings have additional spatially textured modes. Plurimodal tilings include most periodic tilings and feature modes that are typically localised along lines (Fig.~\ref{fig2}h and Appendix)---this type of mode is often encountered in highly symmetric lattices~\cite{Lubensky_review}. The additional modes of oligomodal tilings have particularly interesting kinematics as they can exhibit spatially decaying deformations over a long range (Fig.~\ref{fig2}g). These large-scale deformations, which to the best of our knowledge have not been reported before, offer the prospect of unconventional bulk properties.

\begin{figure*}[t!]
\centering
\includegraphics[width=1\textwidth]{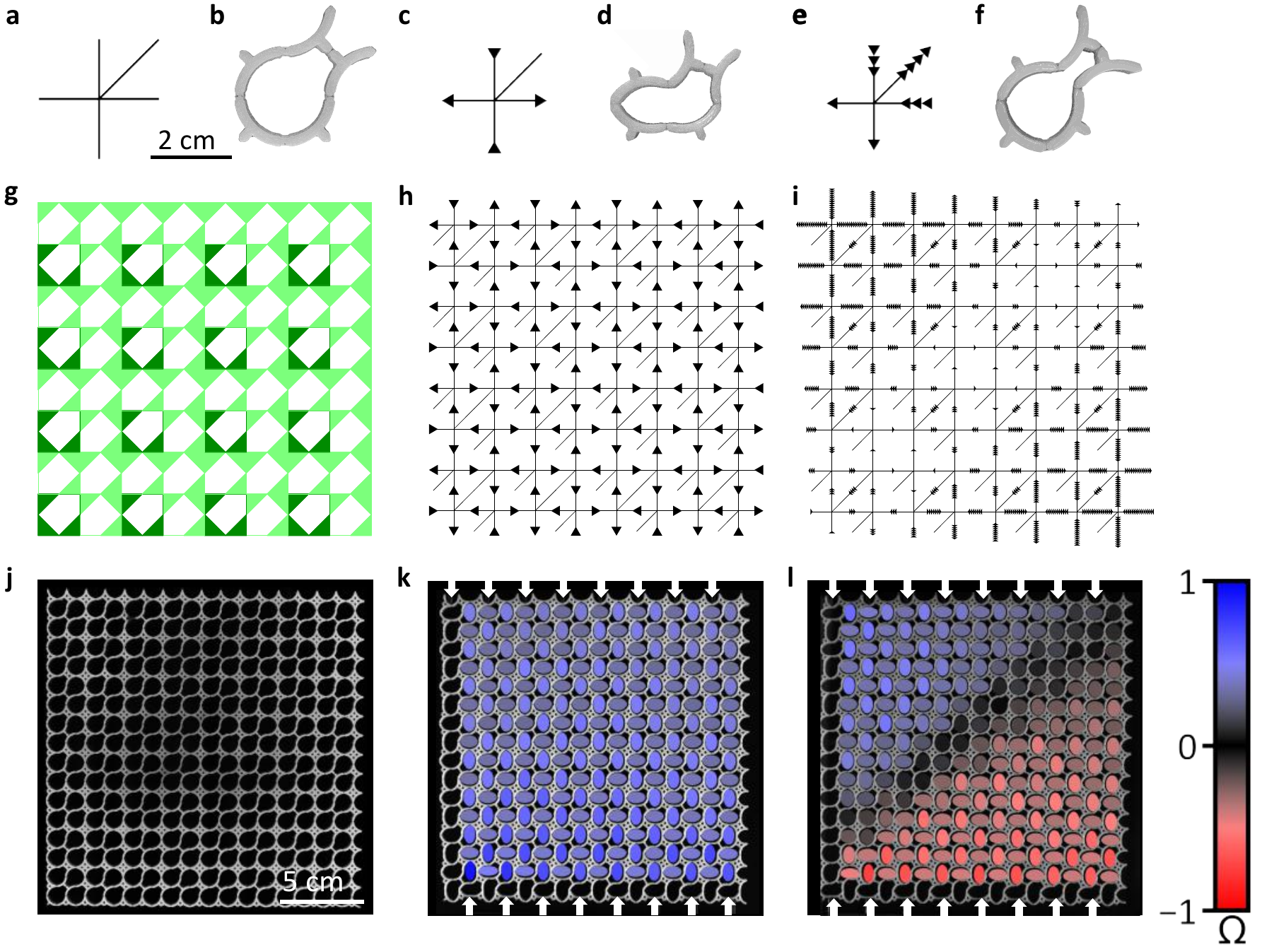}
\caption{\textbf{Oligomodal metamaterials} (a-f) Unit cell of Fig. \ref{fig2}a in vertex representation (ace) and 3D-printed configuration (bdf) in undeformed (ab), deformed in its first mode (cd) and deformed in its second mode (ef) configurations. (g-i) $8 \times 8$ Schematic oligomodal metamaterial design (g), with rotating-squares mode (h) and bi-domain mode (i) visualised in vertex representation. (j-l) $16 \times 16$ Oligomodal metamaterial sample at rest (j), subjected to textured boundary conditions 1 (k) and 2 (l) (white arrows). Coloured ellipses indicate hole polarisation $\Omega$ defined in the Main Text. }
\label{fig3}
\end{figure*}

In the following, we focus on the simplest oligomodal configuration that we uncovered: the periodic tiling of Fig. \ref{fig2}g that can host two bulk modes irrespective of system size. In order to  understand the nature of its soft deformation modes (and in fact also to interpret and classify all three classes of tilings, see Appendix), we introduce a vertex representation that maps out angle deformations onto a directed graph, that is reminiscent of vertex models found in 2D statistical physics~\cite{Bernal_JChemPhys1933,Lieb_PR1967,Baxter_AnnPhys1972} (Appendix). This representation allows for a simple graphical representation of single-cell deformations (Fig. \ref{fig2}a). Each vertex corresponds to a unit cell, each edge corresponds to a hinge, and the number of arrowheads on each edge quantifies the hinge deformation (Fig.~\ref{fig3}ace). In this representation, the geometric constraints can be formulated as follows: only linear combinations of the two vertices shown in Fig. \ref{fig3}ce are allowed. These geometric constraints are propagated from one vertex to its neighbours via the edges through a combinatorial search. Unlike the standard structural analysis computation that we used to predict the number of modes shown in Fig.~\ref{fig2}e (Appendix), the vertex model provides direct insight into the spatial shape of the possible soft deformation modes. Performing such an analysis on the oligomodal tiling shown in Fig.~\ref{fig3}g reveals two distinct spatially extended modes. First, we find the rotating-squares mode (\ref{fig3}h) as expected and consistent with Fig.~\ref{fig2}f. Second, we find a new bulk bi-domain mode, which features a symmetry axis going from South-West to North-East, splitting the deformation into two domains (Fig. \ref{fig3}i) consistent with Fig.~\ref{fig3}g. In the remainder of the paper, we demonstrate that we can selectively actuate these modes---and thus obtain two distinct effective mechanical properties---by either using boundary texture or strain rate dependent viscoelasticity. 

First, we demonstrate that sets of suitably chosen textured boundary conditions under simple uniaxial compression allow to preferentially excite a single deformation mode of choice while frustrating the others. Using multimaterial 3D printing, we create a metamaterial made of $16\times16$ unit cells, following the oligomodal tiling of Fig.~\ref{fig3}g. The 3D printed unit cells are made of rigid curved struts and compliant hinges (Fig.~\ref{fig3}b, see Appendix for details), such that their shapes coincide exactly with the idealised unit cell (Fig.~\ref{fig2}a). Crucially, the 3D printed unit cells have two soft deformation modes that closely mimic that of the idealised unit cell (See Figs.~\ref{fig2}a-\ref{fig3}df). Following the deformation prescribed by the vertex representation (Figs.~\ref{fig3}hi), we then compress our metamaterial using two sets of 8 indenters on each side, that apply the load on every other unit cell and where the bottom set of indenters is either anti-aligned (Fig.~\ref{fig3}k) or aligned (Fig.~\ref{fig3}l) with the top set of indenters. To quantify the deformations of each unit cell, we use standard image tessellation techniques to quantify the flattening $f$ and orientation $\phi$ w.r.t. the horizontal of each pore and calculate the polarisation \cite{Dykstra_JAM2019} $\Omega:=(-1)^{n_x+n_y} f\cos 2\phi$, where $n_x$ ($n_y$) is the unit cell's column (row). Using this protocol, we observe that we can actuate either the rotating-squares mode (Fig. \ref{fig3}k) or the bi-domain mode (Fig. \ref{fig3}l) by changing the location of load application. This example illustrates that the response of oligomodal metamaterials can be selectively actuated by tuning the boundary texture. 
In the following, we build on this result to achieve multi-functionality under a simpler actuation protocol that does not resort to boundary texture.

To achieve this goal, we leave the realm of static responses and harness dissipation~\cite{Dykstra_JAM2019} to tune the dynamical response of our metamaterial. We focus on auxetic behaviour, as it is perhaps the most famous and paradigmatic unusual property of mechanical metamaterials~\cite{Bertoldi_NatRevMat}. We demonstrate that the mechanical response of oligomodal metamaterials can be switched on-demand between negative and positive Poisson's ratios by applying a uniaxial compression at small or large strain rate, respectively. In practice, we use the oligomodal metamaterial design of Fig. \ref{fig3}g, which we rotate by $45^\circ$. With this rotation of the lattice, we expect the domain wall seen in Fig.~\ref{fig3}l to rotate by $45^\circ$, and thus to become perpendicular to the loading axis as well as to provide mirror symmetry about the transverse axis. We expect the presence of this domain wall to induce a larger Poisson's ratio for the bi-domain mode than for the rotating-squares mechanism than---see Appendix.

In order to selectively actuate one mode or the other, we use a combination of stiff material for the rigid parts and soft material for the compliant hinges. In particular, we dress our metamaterial with two types of elastic and viscoelastic hinges (Fig. \ref{fig4}ab)---see Appendix for details about the design, fabrication and calibration. For slow actuation rates, both types of compliant hinges have a very similar torsional stiffness (Appendix). When the actuation rate increases, the viscoelastic hinges stiffen dramatically: they become more than five times stiffer than the elastic hinges (Appendix). Therefore, at slow (fast) loading rates we expect that the central elastic hinge of each supercell will not (will) be actuated hence leading to the rotating-squares (bi-domain) mode, see Fig. \ref{fig4}d (Fig. \ref{fig4}e). We compress the metamaterial slowly at a strain rate $\dot{\varepsilon}=9.3 \cdot 10^{-6} s^{-1}$ (fast at a strain rate $\dot{\varepsilon}=0.11 s^{-1}$). We indeed observe that the metamaterial responds in the rotating-squares (bi-domain) mode, see Fig. \ref{fig4}f (Fig. \ref{fig4}g). The polarisation $\Omega$, overlaid on the experimental pictures, allows to clearly visualise the overall negative polarisation (polarisation split into two domains) for the rotating squares (bi-domain) mode. This result shows that we can use strain rate dependent viscoelasticity to switch between two soft deformation modes. 

Because of the nature of the deformation of the unit cell (Fig. \ref{fig3}a-f), these two modes induce very different Poisson's ratios in the post-buckling regime---see Appendix. Namely, the rotating-squares mode leads to a response with a negative nonlinear Poisson's ratio at strains $\varepsilon>0.04$ while the bi-domain mode induces a positive nonlinear Poisson's ratio (Fig. \ref{fig4}j). Therefore, our metamaterial is multifunctional, hosting several functions that can be selected with a robust actuation protocol: auxetic under slow compression rate and non-auxetic at fast compression rate.

To conclude, we have introduced a novel class of materials, oligomodal metamaterials, that strike a delicate kinematic balance that allows them to exhibit a limited number of deformation modes. These modes can then be selectively and robustly actuated using relatively simple loading protocols such as uniaxial compression, thus providing an example of multifunctional mechanics. We anticipate applications in soft robotics, bio-mimicking systems and shock damping. 

Our approach could be further generalised to other types of unit cells, to other lattices and to three-dimensional tilings. Also, our design space, mostly restricted to periodic tilings of $2\times2$ supercells, could be dramatically augmented for instance by considering periodic tilings of larger supercells or alternative non-periodic tiling strategies constructed via fractal substitution rules and mirroring (See Appendix for an example). 
Finally, data-driven approaches such as machine learning could provide a powerful alternative to better understand the structure-property relationship and to rationally design oligomodal tilings. Additional open questions remain how to effectively design large deformation responses, which variety of multifunctionality can be achieved and how to produce these materials on a large scale effectively. 
\section*{Acknowledgements}
We thank D. Giesen and S. Koot for their skilful technical assistance and R. van Mastrigt, D. Bonn and M. van Hecke for their insightful discussions. C.C. acknowledges funding from the European Research Council grant ERC-StG-Coulais-852587-Extr3Me.

\bibliography{References}


\begin{figure*}[t!]
\centering
 \hspace{0in} \includegraphics[width=\columnwidth]{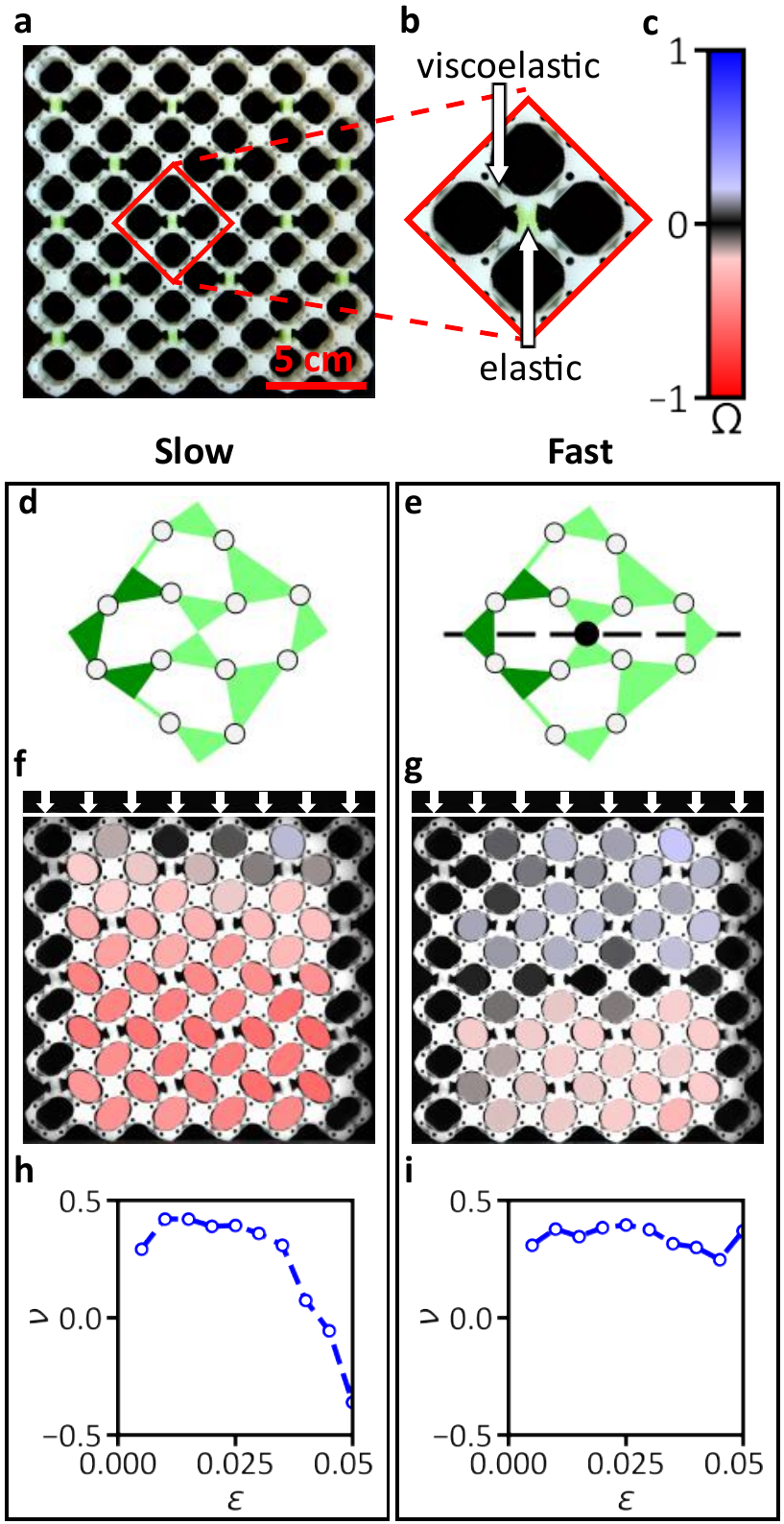}  \\
\caption{\textbf{Multifunctionality mediated by viscoelasticity} (a-b) Oligomodal metamaterial made of rigid (white), elastic (green) and viscoelastic (transparent) material at rest. The tiling design follows Fig.~\ref{fig3}, but is rotated by $45^\circ$. (b) Close-up on a $2\times2$ supercell. (c) Colourbar indicating ellipticity $\Omega$ in (fg). Response of oligomodal metamaterial under slow (dfh) and fast (egi) compression rates. (de) Schematic representation of the counter-rotating mode expected under slow actuation (d) and of the bi-domain mode expected under fast actuation (e). (fg) Snapshot of the metamaterial and overlaid reconstructed pores after a compression $\varepsilon=0.056$ at a strain rate $\dot{\varepsilon} = 9.3 \cdot 10^{-6} s^{-1}$ (f) and $\dot{\varepsilon}=0.11 s^{-1}$ (g), with corresponding nonlinear Poisson's ratio (h) and (i) respectively. The polarisation is defined as $\Omega:=(-1)^{n_y} f\cos 2\phi$, where $n_y$ is the unit cell's row, starting at 0 at the bottom, $f$ is the flattening of the ellipse and $\phi$ is the orientation of the ellipse w.r.t. the first diagonal.}
\label{fig4}
\end{figure*}
\clearpage
\setcounter{equation}{0}
\renewcommand{\theequation}{A\arabic{equation}}%
\setcounter{figure}{0}
\renewcommand{\thefigure}{A\arabic{figure}}%
\setcounter{table}{0}
\renewcommand{\thetable}{A\arabic{table}}%

\section{Appendix}

\subsection{Kinematics of multimode combinatorial metamaterials}
Repeating our bi-mode cell with arbitrary orientations on a square lattice yields a large variety of configurations. To understand the balance between constraints and degrees of freedom, we use the Maxwell-Calladine index theorem~\cite{Lubensky_review}. It provides a relation between the number of states of self-stress $N_{SS}$, the number of mechanisms $N_M$, the dimension $d$, the number of hinges $N$ and the number of constraints $c$,
		\begin{equation}
			dN-c = \frac{d(d+1)}{2}+N_M - N_{SS},
			\label{eq:indexthm}
		\end{equation}
which we can evaluate for arbitrary choices of cell orientations. In a tiling of $n\times n$ cells, each cell has a contribution of $N = 3$ and $c = 7$, with an additional contribution of $N = 4n+1$ and $c = 4n$ from the boundary (Fig. \ref{maxwellcount}a). Constraining displacements to the plane, this yields
		\begin{equation}
			N_M - N_{SS} = 4n - n^2 - 1.
			\label{eq:periarea}
		\end{equation}
 Since the number of modes $N_M$ can not be negative, Eq. (\ref{eq:periarea}) implies a lower bound for the number of states of self-stress of $n^2 - 4n + 1$. Thus, to leading order, the number of states of self-stress is extensive. Eq. (\ref{eq:periarea}) provides a negative lower bound for $N_M$, which hence is poor. In order to obtain more information on the number of modes, one has to go beyond the Maxwell-Calladine index theorem by computing the kernel of the compatibility matrix explicitly~\cite{Lubensky_review}. To obtain the mode scalings displayed in Fig. \ref{fig2}e of the Main Text, we determine this matrix numerically and calculate its rank numerically using a singular value decomposition algorithm. 

\subsection{Vertex model}
\begin{figure}[t!]
\centering
\hspace{0in} \includegraphics[width=\columnwidth]{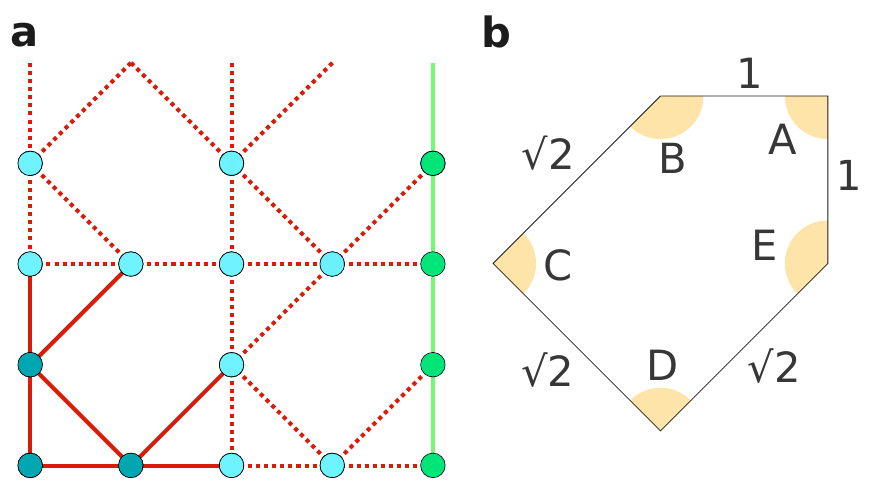}
\\
\caption{\textbf{Kinematics of the unit cell} (a) Counting the constraints and degrees of freedom. Each cell contributes $N=3$ points and $c=7$ constraints. Boundary cells additionally contribute $N=2$ points and $c=2$ constraints, in green. (b) Geometry and parametrisation of the five-bars linkage, which constitutes the functional backbone of the bi-modal unit cell.}
\label{maxwellcount}
\end{figure}

 The approach described above has however a drawback: it yields arbitrary superpositions of the zero modes, potentially providing very little insight on the nature of the involved deformations. To complement this standard method, we therefore introduced a novel approach in the Main Text, based on graphs with directed edges. We now derive this approach in detail, starting from the fully non-linear geometric constraints of the primitive cell. They can be encoded in three trigonometric equations, namely
\begin{equation}
	A+B+C+D+E = 3\pi,
	\label{eq:anglesum}
\end{equation}
\begin{equation}
	1 - \cos(A) = 3 - 2 \cos(C) - 2 \cos(D) + 2 \cos(C+D),
	\label{eq:cosinelaw}
\end{equation}
\begin{equation}
	\sin(D) - \frac{\sin(D+E)}{\sqrt{2}} = \sin(C) - \frac{\sin(C+B)}{\sqrt{2}}, 
	\label{eq:sinelaw}
\end{equation}
where the angles $A$, $B$, $C$, $D$ and $E$ are defined in Fig. \ref{maxwellcount}b.
To construct the directed graph model, we perform a few algebraic manipulations and linearisation. First, we change variables and consider
\begin{multline}
	A = \frac{\pi}{2} + \alpha\quad B = \frac{3\pi}{4} + \beta\quad\\ C = \frac{\pi}{2} + \gamma\quad D = \frac{\pi}{2} + \delta\quad E = \frac{3\pi}{4} + \varepsilon.
	\label{eq:restangles}
\end{multline}
Then, we linearise Eqs. (\ref{eq:cosinelaw}) and (\ref{eq:sinelaw}) in $\alpha$, $\beta$, $\gamma$, $\delta$ and $\varepsilon$. This linearisation around the rest position of our primitive cell yields
\begin{equation}
	\begin{pmatrix}
		\alpha\\
		\delta\\
		\varepsilon
	\end{pmatrix}
	=
	\begin{pmatrix}
		-2 & -2\\
		-1 & -2\\
		 2 & 3
	\end{pmatrix}
	\begin{pmatrix}
		\beta\\
		\gamma
	\end{pmatrix}.
	\label{eq:system}
\end{equation}
As expected from the index theorem, there are two free parameters. 
We then pick a convenient basis 
\begin{equation}
	\begin{pmatrix}
		\beta\\
		\gamma
	\end{pmatrix}
	=
	\chi
	\begin{pmatrix}
		-1\\
		1
	\end{pmatrix}
	+
	\xi
	\begin{pmatrix}
		3\\
		-1
	\end{pmatrix}.
	\label{eq:chixi}
\end{equation}
This equation allows to design Fig. \ref{fig1}c. We draw the cell as a vertex with five edges, one for each hinge, and make use of the fact that only integer coefficients appear in Eq. (\ref{eq:system}) to draw arrows on the edges. Setting $\xi=0$ ($\chi=0$), we draw $n$ arrows on an edge if the corresponding angle is equal to $n\chi$ ($n\xi$). We draw the arrows as incoming for positive angles and outgoing for negative angles. This produces the directed vertex of Fig. \ref{fig2}c (\ref{fig2}d). Any linear combination of those two configurations also produces a compatible vertex. The final ingredient in our model consists in remarking that every hinge is also subject to angle conservation, allowing us to concatenate single-cell graphs to obtain a directed graph describing the full tiling (Fig. \ref{fig2}hi).
\begin{table}[t!]
\centering
\hspace{0in} \includegraphics[width=\columnwidth]{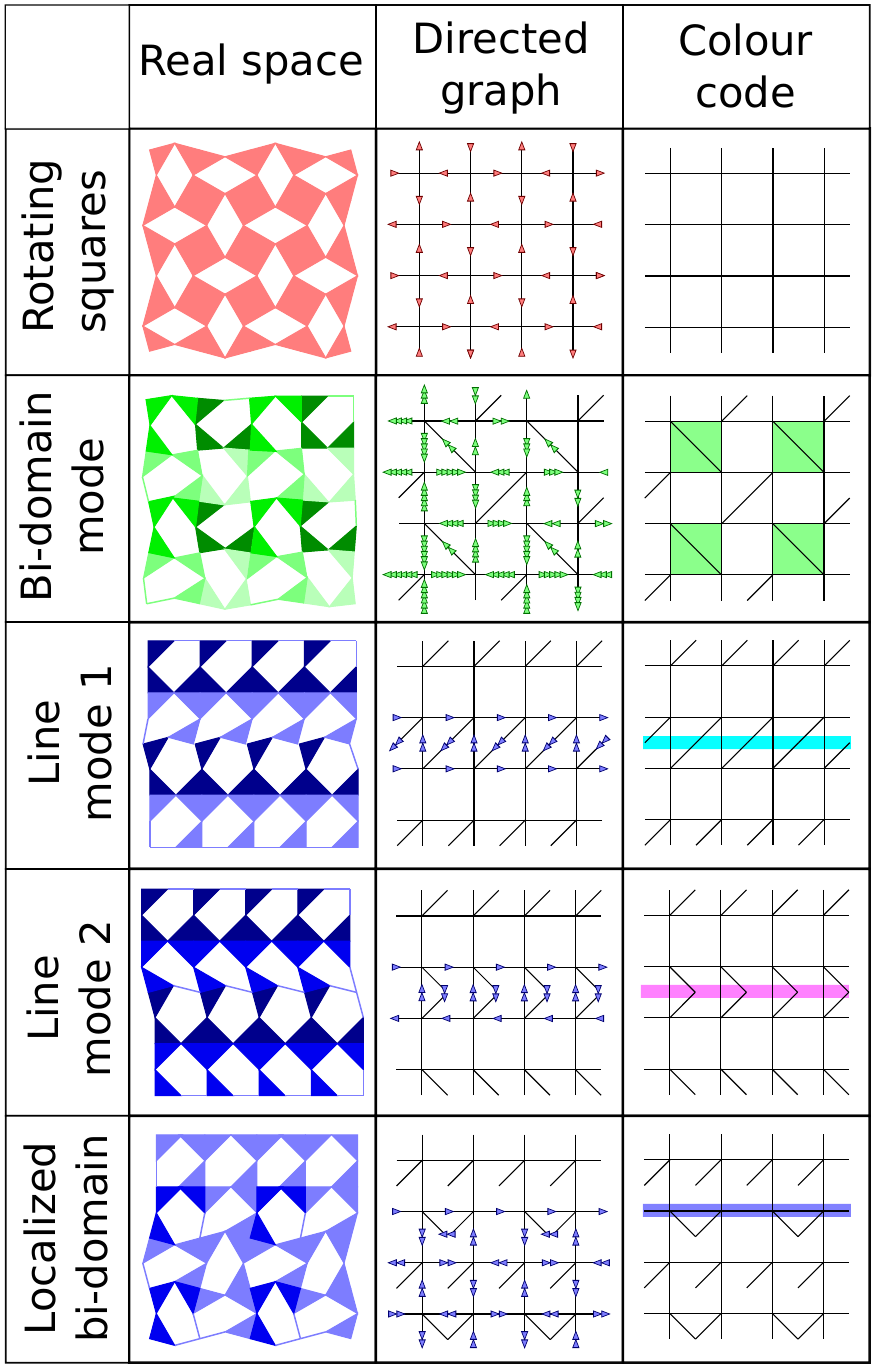}  \\
\caption{\textbf{Table of elementary modes.} The first column depicts the five elementary modes found for 2x2 supercells in real space. The second column represents the associated arrow configurations. The third column is a short-hand notation indicating the presence of a mode in the graph. Since the rotating-squares mechanism (first row) is present in all tilings by construction, we do not associate a colour code to it.}
\label{rosetta}
\end{table}
			
\begin{table}[t!]
\centering
\hspace{0in} \includegraphics[width=\columnwidth]{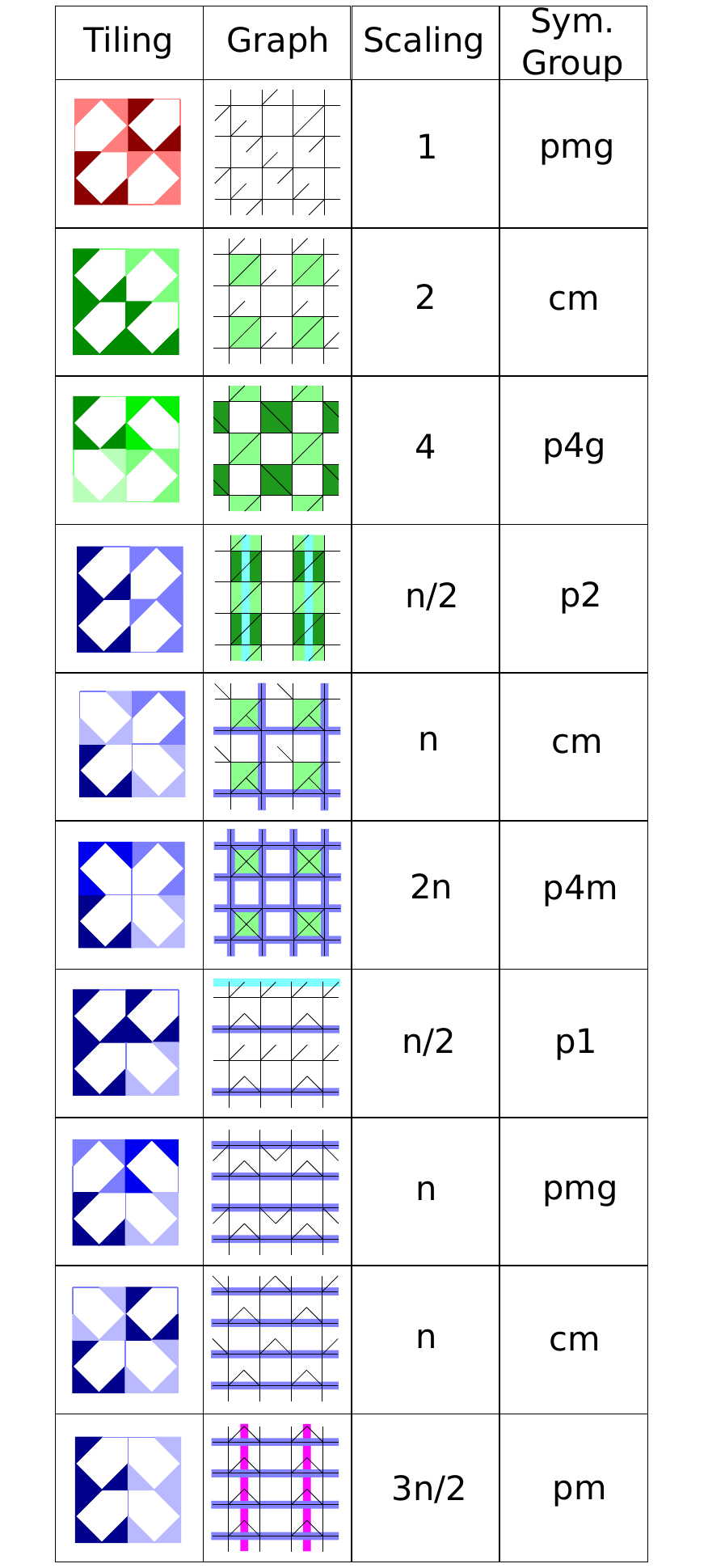}  \\
\caption{\textbf{Table of 2x2 supercells.} The first column lists the ten possible mechanically non-equivalent supercells. The second column lists the associated graphs, with the location of the modes highlighted according to the colour code of column one. The third column gives the leading term of the asymptotic mode scaling as a function of the side length $n$, for a square system. The last column lists the wallpaper group of each tiling.}
\label{2x2table}
\end{table}

Importantly, by an angle-fixing argument, the model could be trivially extended to six, seven and eight edges per vertex by taking our existing vertex and adding a diagonal edge in either of the three available corners. If we then fix the value of one of these diagonal edges, we reduce the problem to the pentagonal case. Since this yields three, four or five independent vertices, it exhausts the number of available modes. In principle, this linearisation procedure is also applicable to other primitive cells, and provides a convenient graphical tool in the cases for which the resulting coefficients are integer.

\subsection{Tiling zoology}
	\label{sec:periodic}
It is useful to describe the actuation of a few fundamental mechanisms in the vertex model framework, in order to understand the global zero modes of the tilings. Five such mechanisms are depicted in Tab.~\ref{rosetta}, with the associated directed graphs. 
	
We restricted our attention to periodic configurations with $2\times 2$ supercells. In practice, only the supercells that are not equivalent are depicted in the first column of Tab.~\ref{2x2table}. We explicitly checked that the vertex graphs of the other 246 tilings are congruent to these ten, if we erase open-ended diagonal edges, which are irrelevant by virtue of having a fixed angle, and up to the presence of surface modes of depth one. In order to see whether tiling symmetries have any bearing on the mode number, shape and localisation, we specify the wallpaper group of each tiling~\cite{Radaelli_Book2011}. Using the colour code of Fig.~\ref{rosetta}, we highlight the locations of the involved mechanisms for each of these configurations in Tab.~\ref{2x2table}, and give the leading term of mode scaling for large system sizes, which can be obtained either by the numerical compatibility matrix approach or by direct mode counting in our graphical method.  The rotating-squares mechanism is not indicated, since its presence is guaranteed for all tilings. Since some graphs with identical symmetries exhibit different mechanical behaviours, we see that the wallpaper group of the tiling does not directly determine the mechanisms it can host.

This analysis based on the vertex model allows to rationalise the rich spectrum of soft deformation modes, comprising three different scalings with system size (Tab.~\ref{2x2table}) and discussed in the Main Text: monomodal tilings with a single mode, irrespective of system size; oligomodal tilings with a constant mode number larger than one; and plurimodal tilings with an increasing number of modes.

We discuss here the plurimodal tilings in more detail. Plurimodal tilings include most periodic tilings and features modes that are typically localised along lines (Tab.~\ref{rosetta}). An insightful way to interpret these modes is to see them as surface modes that survive increases in system size by being mechanically compatible with their neighbouring surface in the bulk. Hence, the translation invariance of periodic systems seems to lead to localised bulk modes that scale linearly with system size (Tab.~\ref{rosetta}). This interpretation can be generalised to different tiling constructions, as we discuss below.

\subsection{Quasiperiodic configurations}
\label{sec:aperio}

\begin{figure*}[t!]
\centering
 \hspace{0in} \includegraphics[width=1\textwidth]{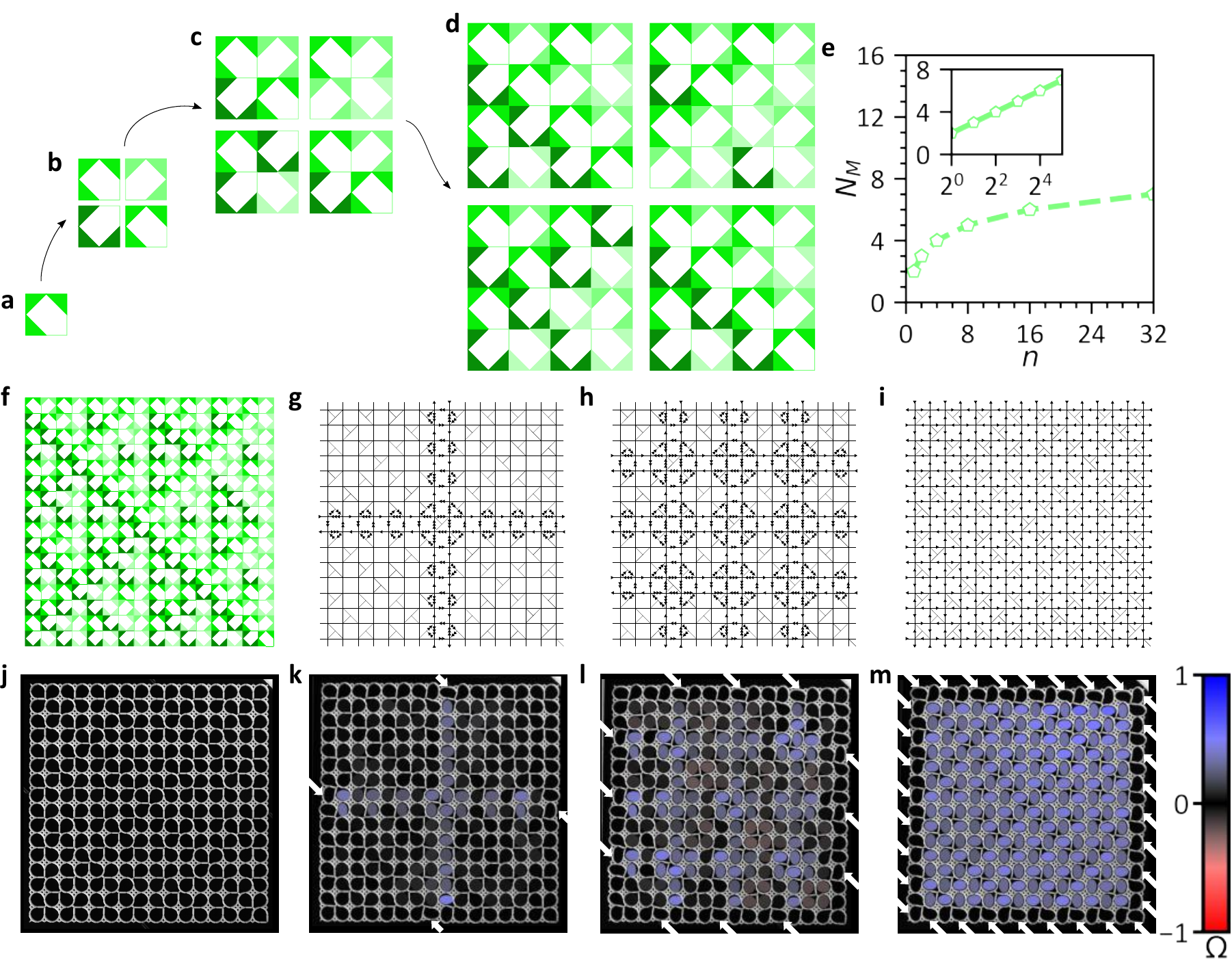}  \\
\caption{\textbf{Construction rule for quasiperiodic tilings.} The tiling in each panel is obtained from the previous one by mirroring it to the left and to the bottom, and translating it in the lower-left diagonal. The first five iterations are shown, (a-d,f) respectively. (e) Logarithmic mode scaling, $N_M$, with system size, $n$. (f-i) $16 \times 16$ quasiperiodic metamaterial in schematic representation (f) and bulk mode vertex model representation: cross mode 1 (g), cross mode 2 (h) and mode with rotating-squares mode (i). (j-m) Corresponding $16 \times 16$ quasiperiodic metamaterial sample at rest (j), subjected to textured boundary conditions 1 (k), 2 (l) and 3 (m) (white arrows). Coloured ellipses indicate hole polarisation $\Omega$.}
\label{appendixfig:quasi}
\end{figure*}

Disordered configurations being trivial does not preclude aperiodic configurations altogether. Indeed, a further class of arrangements can be investigated, that of quasiperiodic configurations. These can be obtained through a copy-and-paste procedure (Fig. \ref{appendixfig:quasi}a-e) that emphasises the coherent matching of surface modes and the resulting hierarchy of bulk modes, seen in Fig. \ref{appendixfig:quasi}g-i. The procedure used for the structure we investigated experimentally goes as follows: mirror the full tiling to the right and to the bottom, and translate it to fill the resulting gap. Alternatively, it can be seen as a substitution rule, in which each primitive cell is replaced by a particular group of four cells. 

As in the case of plurimodal tilings, surface modes can survive the growth process and yield mechanical bulk modes, due to their inherent compatibility. These modes are intriguing: they scale logarithmically with the system size (Fig. \ref{appendixfig:quasi}e) and exhibit a peculiar spatial structure, in the form of cross-shapes (Fig.~\ref{appendixfig:quasi}gh). We illustrate these modes experimentally on a $16\times 16$ metamaterial with three bulk modes---the rotating-squares mode and two cross-shaped bulk modes (as in Fig. \ref{appendixfig:quasi}g-i)---and three modes localised at the edges. We perform three distinct experiments with three sets of textured boundaries that allow us to successfully actuate the three bulk modes upon compression. While multiple indenters allow us to obtain the rotating-squares mode (Fig. \ref{appendixfig:quasi}m), only a few allow to obtain a cross-shaped mode (Fig. \ref{appendixfig:quasi}k) or a mode with three cross-shaped modes (Fig. \ref{appendixfig:quasi}l). It would also be possible to obtain a deformation with two cross-shapes, equivalent to subtracting Fig. \ref{appendixfig:quasi}g from Fig. \ref{appendixfig:quasi}h. 

Not only is this method applicable to quasiperiodic structures but it could form the base of an alternative method to design periodic oligomodal metamaterials. By mirroring a metamaterial with a surface mode, the surface mode becomes inherently compatible. After mirroring, we then obtain a supercell, which is inherently suitable for a periodic tiling. This can clearly be seen by noting that the modes of Fig.~\ref{appendixfig:quasi}fg can both be tiled in a periodic manner: the bottom side is compatible with the top side and the left side is compatible with the right side.

This example illustrates that our combinatorial design space is potentially very large. Quasiperiodic tilings could also be further investigated through choices of $2\times2$, $3\times3$, ... substitution rules, some of which yield periodic tilings. This is another landscape of non-trivial combinatorial metamaterials, whose systematic study we leave for future works. 

\subsection{Finite mechanisms and infinitesimal modes}

We discuss here the nature of the modes determined above. Two classes of modes can exist~\cite{Lubensky_review}: (i) the finite mechanisms, in which the zero-energy deformations can persist in the nonlinear regime and lead to large rotations without stretching or compressing any bars; (ii) infinitesimal modes, in which the deformations are only zero-energy in the linear range. Further deformations in the nonlinear regime typically cost elastic energy through length-change of the rods. 

A visual inspection of the kinematics of the elementary modes in Tab.~\ref{rosetta} reveals that only the rotating-squares mode is finite. All the other modes require length-change of the rods to be extended into the nonlinear regime and as such are infinitesimal. 

\subsection{Sample design and fabrication}
The 16$\times$16 samples of Fig. \ref{fig3} are designed following the tilings shown in Fig.~\ref{fig2}. The samples' dimensions are $192 \times 192\times 4$ mm.
The 6$\times$6 sample of Fig. \ref{fig4} is designed following the tiling shown in Fig.~\ref{fig2}ab, rotated by $45^o$. The sample's dimensions are $180 \times 180\times 20$ mm.

\begin{figure}[t!]
\centering
 \hspace{0in} \includegraphics[width=\columnwidth]{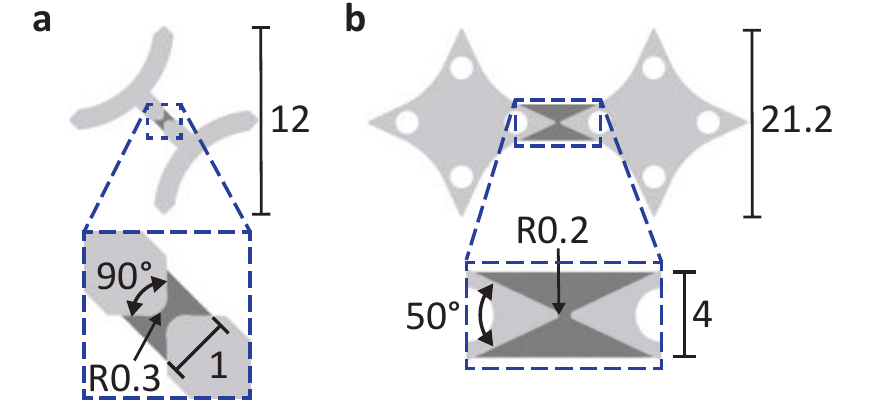}  \\
\caption{\textbf{Compliant hinge geometry} (a) Hinge geometry of Fig. \ref{fig3}. (b) Hinge geometry of Fig. \ref{fig4}. The light grey (dark grey) material is rigid (compliant). All units are in mm unless otherwise specified.}
\label{fig7}
\end{figure}

For these two types of samples, special care has been devoted to the design of the details of the unit cells: imperfect compliant hinges, experiencing shear and tension instead of pure bending/torsion, can affect the linear and nonlinear deformation of mechanisms significantly~\cite{Coulais_NatPhys2018}. For the samples of Fig.~\ref{fig3} and Fig. \ref{fig4}, we negate these effects in two ways. First of all, the central unit cells are made from a much more rigid material than the elastic hinges, such that the central unit cells do not deform during loading. Next, we use tapered rigid parts as seen in Fig.~\ref{fig7}. A central narrow section between the rigid sections greatly increases resistance to shear and tension while offering only a negligible resistance to bending.

When the actuation is aligned with the primitive lattice, the bi-domain mode leads to shear-compression coupling. However, such a mode in a $45^\circ$ orientation will lead to a positive Poisson's ratio, which is markedly different from the rotating-squares modes, that has a negative Poisson's ratio. Therefore, we choose to rotate the sample by $45^\circ$ in Fig. \ref{fig4}. Without imperfection, the configuration in Fig. \ref{fig4} would have an inherent buckling preference towards the rotating-squares mode of Fig. \ref{fig4}e. However, this preference is tuned by designing an imperfection~\cite{Coulais_Nature2018} favouring the bi-domain mode of Fig. \ref{fig4}d, which has been designed using nonlinear static finite element methods (Abaqus). Furthermore, as the samples are tested horizontally, the bottom of the sample features small pockets with 5mm steel ball bearings to reduce the effects of friction with the bottom surface.

We produce the samples in Fig.~\ref{fig3} using additive manufacturing using a Stratasys Objet500 Connex3 3D printer. The hinges are produced from a highly viscoelastic rubber-like PolyJet Photopolymer (Stratasys Agilus 30), which is co-cured with the central parts, which are made from a much stiffer Photopolymer (Stratasys VeroWhitePlus). The square hinges and central parts of the sample of Fig. \ref{fig4} are produced in the same way. However, the triangular hinges of the sample of Fig. \ref{fig4} are cast from an elastic 2-component silicon-based rubber (Zhermack Elite Double 32), which are then glued to the sample using Wacker Elastosil E43 glue.

\subsection{Experimental methods}
We compress the periodic bi-mode sample in Fig. \ref{fig3}j-l of the Main Text using a uniaxial testing device (Instron 5943 with a 500 N load cell), with laser cut plexiglas indenters, denoted by the white arrows in the figure, at a strain rate of $\dot{\varepsilon}=1.30 \cdot 10^{-3} s^{-1}$ to a strain of $\varepsilon=0.026$.  The sample is positioned horizontally to enhance overall mechanical stability and prevent out-of-plane buckling. The quasiperiodic sample in Fig. \ref{appendixfig:quasi}j-m is produced and tested in a similar way. However, the sample is compressed instead in the direction from North-West (North and West side) to South-East (South and East side).

The sample in Fig.~\ref{fig4} of the Main Text is compressed using the same uniaxial testing machine with straight surfaces on the top and bottom at various strain rates. The sample is also positioned horizontally to enhance overall mechanical stability prevent out-of-plane buckling. For Fig. \ref{fig3} (Fig. \ref{fig4}), the tests are recorded using a high resolution $3858 \times 2748$ ($2048 \times 2048$) monochrome CMOS camera Basler acA3800-14um with Fujinon 75 mm lens (acA2040-90um with Fujinon 50 mm lens), inducing a spatial resolution of $0.07$ mm ($0.1$ mm). Using standard image tesselation techniques, we extract and track the four central holes of the stiff sections, which we use to calculate the position and ellipticity of all the pores, whose reconstruction has been overlaid in Fig. \ref{fig3} (Fig. \ref{fig4}). For Fig. \ref{fig4}, we track the average horizontal (vertical) distance between the centres of ellipses on the sides (top and bottom) to compute the width (height) of the sample, from which we extract the compressive axial strain $\varepsilon_\textbf{axial}$ and the strain transverse to the loading direction $\varepsilon_\textbf{transverse}$. We use these quantities to compute the nonlinear Poisson's ratio $\nu := -\partial \varepsilon_\textbf{transverse} / \partial \varepsilon_\textbf{axial}$ averaged over applied strain deformation steps of $\delta\varepsilon_\textbf{axial}=0.005$.


\subsection{Hinge material properties}
We measure the mechanical properties of a single compliant hinge of the sample in Fig. \ref{fig4} using the same uniaxial testing device. We perform three different experiments: bending, tension and shear.
For bending, we pull on the hinge up to 0.30 rad deflection at rates of 0.002 and 20 mm/s. For tension and shear, we pull on the hinge up to 1.0 mm deflection at rates of 0.002 and 20 mm/s.
We report the average of the measured stiffnesses in Tab.~\ref{tab1}.

We observe that the stiffnesses at long timescales are similar for both types of hinges. However, at short time scales, the stiffnesses of the elastic (Elite Double 32) hinges increases only by approximately 20\%-30\% compared to long timescales, while the stiffnesses of the viscoelastic (Agilus 30) hinges can increase by more than 1300\%. This difference in loading-rate dependency allows us to obtain the switchable response observed in Fig. \ref{fig4}.

\begin{table}[h!]
    \centering
    \begin{tabular}{|c c | c c | c c |}
    \hline
        \multicolumn{2}{|c|}{\multirow{2}{*}{\textbf{Stiffness}}}& \multicolumn{2}{c |}{\textbf{Elite Double 32}} & \multicolumn{2}{c|}{\textbf{Agilus 30}}   \\
          \multicolumn{2}{|c|}{} & Fast & Slow & Fast & Slow \\
          \hline
         Bending & [Nmm/rad] & 24  &20   & 120 & 20  \\
         & & $\pm$ 1 & $\pm$ 2 &  $\pm$ 25  & $\pm$ 2  \\
         Tension & [N/mm] & 31  & 24 &  630 & 44 \\
         & & $\pm$ 2 & $\pm$ 1 & $\pm$ 150  & $\pm$ 10 \\
         Shear & [N/mm] & 10 & 8.1 & 190 & 18   \\
         & & $\pm$ 1 & $\pm$ 0.4 & $\pm$ 30  & $\pm$ 3 \\
         \hline
    \end{tabular}
    \caption{\textbf{Mechanical properties of compliant hinges of Fig. \ref{fig4}} Bending, tensile and shear stiffness under fast (20 mm/s) and slow (0.002 mm/s) loading rate.}
    \label{tab1}
\end{table}

\subsection{Poisson's ratio of the oligomodal tiling with the bi-domain mode}
The metamaterial in Fig.~\ref{fig4} has two soft deformation modes. When compressed via flat plates, we expect such a metamaterial to undergo a buckling instability, beyond which either the rotating-squares mode or the bi-domain mode will be actuated. We present in this section the derivation of the Poisson's ratio in the prebuckling phase as well as in the post-buckling phase for either of these two modes.

\begin{figure}[t!]
\centering
\hspace{0in} \includegraphics[width=\columnwidth]{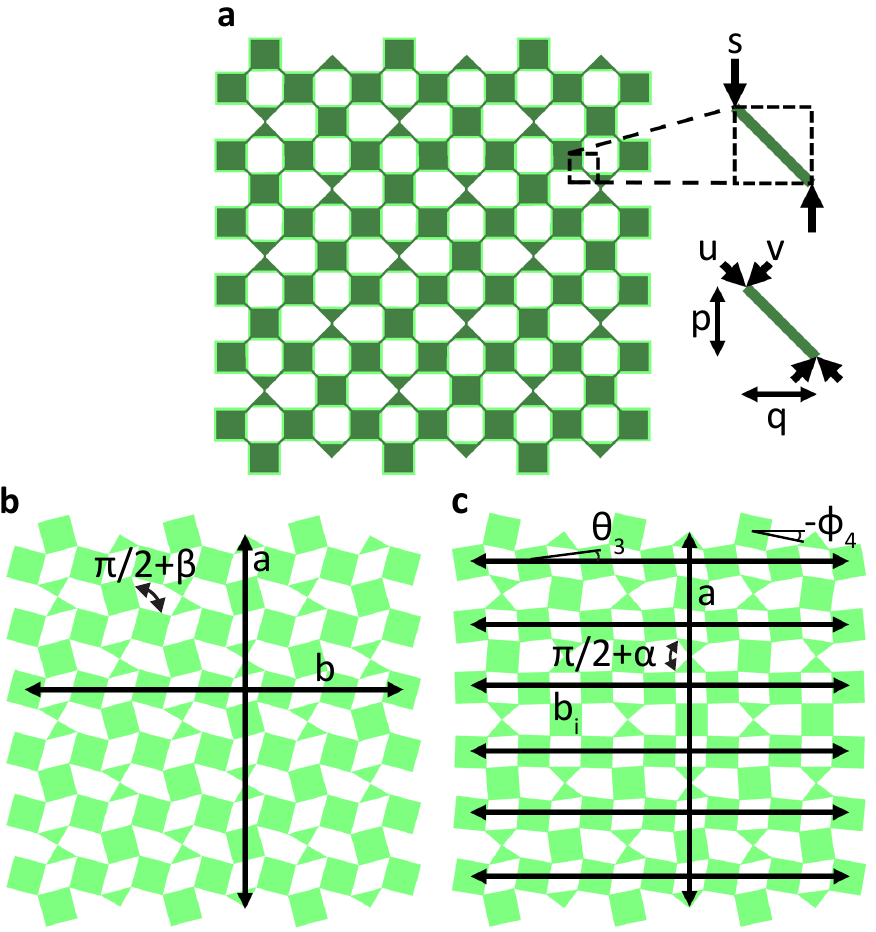}
\\
\caption{\textbf{Poisson's ratio estimation mechanisms} (a) Linear compression model with compression and shear of the hinges for the estimation of the Poisson' ratio in the prebuckling regime. (b) Rotating-squares mechanism. (c) Bi-domain mechanism. $p$ and $a$ ($q$ and $b$) indicate the reference length along (transverse to ) compression. The number of unit cells in horizontal ($n_x$) and vertical ($n_y$) position are both 6.}
\label{FigPoisson}
\end{figure}

The reason why we observe a prebuckling regime over a finite range of strain is the fact that the hinges are not pure torsional hinges---they can compress or elongate and shear as is visualised in Fig. \ref{FigPoisson}a. Hence, to predict the Poisson's ratio in the prebuckling regime, one necessarily has to consider the shear and  extension stiffnesses of these hinges.

Let $s$ be the applied vertical deflection of the hinge, $(u_x,u_y)$ $\left( (v_x,v_y) \right)$ the hinge extensional (shear) deflection and $p$ ($q$) the hinge vertical (horizontal) length respectively. The Poisson's ratio in the linear prebuckling regime $\nu_\textrm{pre}$ can then be calculated as
\begin{equation}
    \nu_\textrm{pre}= -\frac{u_x+v_x}{u_y+v_y}=\frac{k_T-k_S}{k_T+k_S},
\end{equation}
with $k_T$ ($k_S$) the hinge extensional (shear) stiffness. For slow loading---with $k_T = 44\,\pm 10$ N/mm and $k_S = 18\, \pm 3$ N/mm in Tab. \ref{tab1}---, the expected Poisson's ratio is $\nu_\textrm{pre}=0.4 \, \pm 0.2$. This estimation is in quantitative agreement with the Poisson's ratio of the metamaterial measured in the prebuckling regime (Fig. \ref{fig4}hi of the Main Text).

In turn, if we neglect shear and extension of the hinges, we can also estimate the nonlinear post-buckling Poisson's ratio using trigonometry of the underlying mechanisms of Fig \ref{FigPoisson}.
The nonlinear Poisson's ratio is defined as
$\nu:=-\partial b/\partial a (a/b)$, where 
the tiling's horizontal dimension is $b$ and its vertical dimension is $a$.

For the rotating-squares mechanism in Fig. \ref{FigPoisson}b, the tiling's horizontal dimension is $b_{RS}=n_x \cos(\beta/2)$ and its vertical dimension is $a_{RS}=n_y \cos(\beta/2)$. Therefore, we find 
\begin{equation}
    \nu_{RS}=-1.
    \label{eq:nu_RS2}
\end{equation}
For the bi-domain mechanism in Fig. \ref{FigPoisson}c, we find 
\begin{multline}
    a_{BD} = 2 \, \Sigma_{i=1}^{n_y/2} \,\Big(     \frac{1}{2}\cos \phi_{i}  -  \frac{1}{2}\sin \phi_{i}  + \cos \theta_{i}   -  \\
    \sin \theta_{i} +     \frac{1}{2}\cos \phi_{i+1}   -  \frac{1}{2}\sin \phi_{i+1} \Big)
    \label{eq:nu_BD3}
\end{multline}
and

\begin{equation}
    b_{BD,i} = \frac{n_x}{2} \, \left( \cos \phi_{i} + 2 \cos \theta_i + \cos \phi_{i+1} \right),
    \label{eq:nu_BD4}
\end{equation}
where $\phi_i = \frac{\left(1-i\right)}{2} \alpha   $ and $\theta_i = \left(\frac{i-1}{2}+\frac{1}{4}\right) \alpha$. Since $\phi_i$ and $\theta_i$ depend on $i$, the lattice does not experience uniform deformation. We therefore define a local Poisson's ratio as
\begin{equation}
    \nu_{BD,i}= -\frac{\partial b_{BDi,}}{\partial \alpha } \left(\frac{\partial a_{BD}}{\partial \alpha }\right)^{-1} \frac{a_{BD}}{b_{BD,i}}.  
    \label{eq:nu_BD5}
\end{equation}
A Taylor series expansion allows us to compute the local Poisson's ratio as $\nu_{BD,1} = -0.12 + \mathcal{O}(\alpha^2)$ at the center of the lattice and as $\nu_{BD,3} = -2.1 + \mathcal{O}(\alpha^2)$ at the top and bottom of the lattice. 
In practice, the bi-domain mode is infinitesimal, which means that it can not undergo large deformations. This implies that large transverse deformations on the top and bottom sides of the lattice are geometrically frustrated and that the hinges of the lattice also undergo shear and extension. As a result, we expect the overall Poisson's ratio to be closer to the lesser deformed $\nu_{BD,1}\simeq -0.12$ at the center of the lattice.

In summary, since $\nu_\textrm{pre}=0.4$ and $\nu_{RS}=-1$, we can expect to observe a marked transition between a positive Poisson's ratio in the prebuckling regime to an increasingly decreasing and negative Poisson's ratio in post-buckling for the rotating-squares mechanism. 
In contrast, since $\nu_\textrm{pre}=0.4$ and $\nu_{BD}\simeq -0.12$, we expect a much smaller decrease in Poisson's ratio for the bi-domain mode at relatively small post-buckling deflections. These theoretical results are consistent with what we observe experimentally in Fig. \ref{fig4}hi of the Main Text. Furthermore, the relation between the hinge extensional and shear stiffness suggests that the pre-buckling Poisson's ratio could be tailored, offering an increased flexibility to design oligofunctional materials.
\end{document}